\documentclass[aps,draft,showpacs]{revtex4}

\usepackage{amsfonts}
\usepackage{amsmath}
\usepackage{amssymb}
\usepackage{graphicx}

\begin{document}

\title[The $K-$way negativities]{Quantum coherences, $K-$way negativities
and multipartite entanglement}
\author{S. Shelly Sharma}
\email{shelly@uel.br}
\affiliation{Depto. de F\'{\i}sica, Universidade Estadual de Londrina, Londrina
86051-990, PR Brazil }
\author{N. K. Sharma}
\email{nsharma@uel.br}
\affiliation{Depto. de Matem\'{a}tica, Universidade Estadual de Londrina, Londrina
86051-990 PR, Brazil }
\thanks{}

\begin{abstract}
A characterization of multipartite quantum states having $N$ subsystems,
based on negativities of matrices obtained by selective partial
transposition of state operator, is proposed. The $K-$way partial transpose
with respect to a subsystem is constructed by imposing constraints involving
the states of $K$ subsystems of multipartite composite system. The $K-$way
negativity, defined as the negativity of $K-$way partial transpose,
quantifies the $K-$way coherences of the composite system. For an N-partite
system the fraction of $K-$way negativity ($2\leq K\leq N$), contributing to
global negativity, is obtained. The entanglement measures for a given state $%
\widehat{\rho }$ are identified as the partial $K-$way negativities of the
corresponding canonical state.
\end{abstract}

\maketitle

Positive partial transpose (PPT), first introduced by Peres \cite{pere96},
is the most widely used separability criterion for quantum states. It has
been shown to be a necessary and sufficient condition \cite{horo96} for the
separability of qubit-qubit and qubit-qutrit systems. For higher dimensional
systems, positive partial transpose is a necessary condition \cite%
{horo197,horo297}. Negativity \cite{zycz98,vida02} based on Peres Horodecki
PPT criterion has been shown to be an entanglement monotone \cite%
{vida02,eise01}. In a multipartite quantum system composed of N subsystems,
a single subsystem may be entangled to (N-1)systems in distinctly different
ways. For example, in a three qubit system ($ABC$), the subsystem $A$ can
have genuine tripartite entanglement, W- like entanglement, as well as
bipartite entanglement with subsystem $B$ or $C$ alone. The negativity of
partial transpose of state operator $\widehat{\rho }^{ABC}$ with respect to $%
A,$ may, thus have distinct contributions that can be related to genuine
tripartite or bipartite entanglement. The bipartite entanglement, may in
turn be for the pair $AB$, the pair $AC$, or both the pairs. In a recent
article \cite{shel07}, we have discussed the entanglement of three qubit
states using $2-$way and $3-$way negativities. In this article, a
characterization of multipartite quantum states having $N$ subsystems, based
on negativities of matrices obtained by selective partial transposition of
state operator, is proposed. The $K-$way partial transpose with respect to a
subsystem is constructed by imposing constraints involving the states of \ $K
$ subsystems of multipartite composite system. The $K-$way negativity ($%
2\leq K\leq N)$, defined as the negativity of $K-$way partial transpose,
quantifies the $K-$way coherences of the composite system. The underlying
idea of selective transposition to construct a $K-$way partial transpose
with respect to a subsystem, first presented in ref. \cite{shel06}, shifts
the focus from $K-$subsystems to $K-$way coherences of the composite system.
By $K-$way coherences, we mean the quantum correlations responsible for GHZ
state like entanglement of a K-partite system. For an N-partite entangled
state, the negativity of global partial transpose is found to contain
contributions from $K-$way partial transposes ($2\leq K\leq N$).
Entanglement is invariant under local unitary rotations, whereas, coherences
are not so. The elements in the set of states obtained by performing
entanglement conserving local operations on $\widehat{\rho }$ differ from
each other by the number of local basis product states and the number of
variables required to write the state. In addition, the states in the set
differ from each other by the $K-$way negativities characterizing the
states. A pure state $\widehat{\rho }=\left\vert \Psi \right\rangle
\left\langle \Psi \right\vert $ may be mapped by local entanglement
conserving operations to an operator $\widehat{\rho }_{c}=\left\vert \Psi
\right\rangle _{c\,c}\left\langle \Psi \right\vert $ such that the canonical
state $\left\vert \Psi \right\rangle _{c}$ is a linear combination of
minimum number of local basis product states (LBPS) \cite{cart99}.  The
entanglement measures for a given state $\widehat{\rho }$ are identified as
the partial $K-$way negativities of the corresponding canonical state $%
\widehat{\rho }_{c}$. A simple method to construct a state canonical to a
given three qubit pure state has been given by Acin et al \cite{acin00}.
While GHZ like N$-$ partite entanglement of a composite system is generated
by N$-$way coherences, N$-$partite entanglement in general can be present
due to $K-$way $\left( 2\leq K<N\right) $ coherences as well. On the other
hand, a system having $K-$partite entanglement may not have $K-$way
coherences at all, as is the case of states having bound entanglement.

In section I, we define the global negativity, the $K-$way partial transpose
of an $N-$partite state and the $K-$way negativity. Decomposition of $K-$way
negativity into contributions from different components of a subsystem is
discussed in section II. Section III deals with the contributions of $K-$way
negativities to the global negativity. In section IV, the effect of local
operations on coherences is discussed. A brief discussion on the connection
between number of eigenvalues of  N-way partial transpose and the canonical
form, in the context of N-qubit state, is given in section V.  Analytical
expressions for entanglement measures of a special set of four parameter two
qubit one qutrit state in Schmidt form are given in section VI, to
illustrate the usage.

\section{The Global and K-way Negativity of N-partite system}

The Hilbert space, $C^{d}=C^{d_{1}}\otimes C^{d_{2}}\otimes ...\otimes
C^{d_{N}}$, associated with a quantum system composed of $N$ sub-systems, is
spanned by basis vectors of the form $\left\vert
i_{1}i_{2}...i_{N}\right\rangle ,$ where $i_{m}=0$ to $\left( d_{m}-1\right)
,$ $d_{m}$ being the dimension of Hilbert space associated with $m^{th}$
sub-system. The state operator for a general N-partite state is 
\begin{equation}
\widehat{\rho }=\sum_{\substack{ i_{1}i_{2}...i_{N},  \\ %
j_{1},j_{2},...j_{N} }}\left\langle i_{1}i_{2}...i_{N}\right\vert \widehat{%
\rho }\left\vert j_{1}j_{2}...j_{N}\right\rangle \left\vert
i_{1}i_{2}...i_{N}\right\rangle \left\langle j_{1}j_{2}...j_{N}\right\vert .
\label{1}
\end{equation}%
The global partial transpose of $\widehat{\rho }_{G}^{T_{p}}$ with respect
to sub-system $p$ is obtained from the matrix $\rho $ by imposing the
condition 
\begin{equation}
\left\langle i_{1}i_{2}...i_{N}\right\vert \widehat{\rho }%
_{G}^{T_{p}}\left\vert j_{1}j_{2}...j_{N}\right\rangle =\left\langle
i_{1}i_{2}...i_{p-1},j_{p},i_{p+1},...i_{N}\right\vert \widehat{\rho }%
\left\vert j_{1}j_{2}...j_{p-1},i_{p},j_{p+1},...i_{N}\right\rangle \text{.}
\label{2}
\end{equation}%
The partial transpose $\widehat{\rho }_{G}^{T_{p}}$ of a state having free
entanglement is non positive. Global Negativity, defined as 
\begin{equation}
N_{G}^{p}=\frac{1}{d_{p}-1}\left( \left\Vert \rho _{G}^{T_{p}}\right\Vert
_{1}-1\right) ,  \label{3}
\end{equation}
measures the entanglement of subsystem $p$ with its complement in a
bipartite split of the composite system. Here $\left\Vert \rho \right\Vert
_{1}$ is the trace norm of $\rho $. Global negativity vanishes on PPT-states
and is equal to the entropy of entanglement on maximally entangled states.

A given matrix element $\left\langle i_{1}i_{2}...i_{N}\right\vert \widehat{%
\rho }\left\vert j_{1}j_{2}...j_{N}\right\rangle $ is characterized by a
number $\sum\limits_{m=1}^{N}(1-\delta _{i_{m},j_{m}})=K$, where $\delta
_{i_{m},j_{m}}=1$ for $i_{m}=j_{m}$, and $\delta _{i_{m},j_{m}}=0$ for $%
i_{m}\neq j_{m}$. In other words, the total number of subsystems in bra
vector in a state different from that in the ket vector in a matrix element $%
\left\langle i_{1}i_{2}...i_{N}\right\vert \widehat{\rho }\left\vert
j_{1}j_{2}...j_{N}\right\rangle $ is equal to $K$. The\ $K-$way partial
transpose ($2\leq K$ $\leq N$) of\ N-partite state $\widehat{\rho }$ with
respect to subsystem $p$ is obtained from matrix $\rho $ by applying the
following constraints: 
\begin{eqnarray}
\left\langle i_{1}i_{2}...i_{N}\right\vert \widehat{\rho }%
_{K}^{T_{p}}\left\vert j_{1}j_{2}...j_{N}\right\rangle  &=&\left\langle
i_{1}i_{2}...i_{p-1},j_{p},i_{p+1},...i_{N}\right\vert \widehat{\rho }%
\left\vert j_{1}j_{2}...j_{p-1},i_{p},j_{p+1},...j_{N}\right\rangle ,\quad  
\notag \\
\text{if}\quad \sum\limits_{m=1}^{N}(1-\delta _{i_{m},j_{m}}) &=&K,\qquad 
\text{and}  \notag \\
\text{ }\left\langle i_{1}i_{2}...i_{N}\right\vert \widehat{\rho }%
_{K}^{T_{p}}\left\vert j_{1}j_{2}...j_{N}\right\rangle  &=&\left\langle
i_{1}i_{2}...i_{N}\right\vert \widehat{\rho }\left\vert
j_{1}j_{2}...j_{N}\right\rangle \quad   \notag \\
\text{if}\quad \sum\limits_{m=1}^{N}(1-\delta _{i_{m},j_{m}}) &\neq &K.
\label{4}
\end{eqnarray}%
The $K-$way negativity \cite{shel06, shel07} calculated from $K-$way partial
transpose of matrix $\rho $ with respect to subsystem $p$, is defined as 
\begin{equation}
N_{K}^{p}=\frac{1}{d_{p}-1}\left( \left\Vert \rho _{K}^{T_{p}}\right\Vert
_{1}-1\right) .  \label{7}
\end{equation}%
Using the definition of trace norm and $tr(\rho _{K}^{T_{p}})=1$, we get%
\begin{equation}
N_{K}^{p}=\frac{2}{d_{p}-1}\sum_{i}\left\vert \lambda _{i}^{K-}\right\vert ,
\label{8}
\end{equation}%
$\lambda _{i}^{K-}$ being the negative eigenvalues of matrix $\rho
_{K}^{T_{p}}$. The negativity $N_{K}^{p}$ depends on $K-$way coherences and
is a measure of all possible types of entanglement attributed to $K-$ way
coherences. Intuitively, for a system to have pure $N-$partite entanglement,
it is necessary that $N-$way coherences are non-zero. On the other hand, $N-$%
partite entanglement can be generated by $(N-1)-$ way coherences, as well.
For a three qubit system, maximally entangled tripartite GHZ state is an
example of genuine tripartite entanglement involving $3-$way coherences. For
maximally entangled three qubit GHZ state the global negativity $%
N_{G}^{p}=N_{3}^{p}=1$. Maximally entangled W-state is a manifestation of
tripartite entanglement due to $2-$way coherences having $%
N_{G}^{p}=N_{2}^{p}=0.94$ and $N_{3}^{p}=0$. For pure states the free
entanglement of a subsystem is completely determined by the global
Negativity and the hierarchy of negativities $N_{K}^{p}$ ($K=2,...N),$
calculated from $\rho _{K}^{T_{p}}$ associated with the $p^{th}$ sub-system.
For N partite system, $N_{2}^{p}$, $N_{3}^{p}$,...,$N_{N}^{p}$ and $N_{G}^{p}
$ ( $p=1$ to $N$) quantify the coherences characterizing bipartite splits of
the system having $p^{th}$ subsystem as one part.

\section{ How do different subsystems contribute to the global partial
transpose?}

The negativity of partial transpose of a two qubit state 
\begin{equation}
\widehat{\rho }^{AB}=\sum\limits_{i_{1}i_{2}j_{1}j_{2}}\left\langle
i_{1}i_{2}\right\vert \widehat{\rho }^{AB}\left\vert j_{1}j_{2}\right\rangle
\left\vert i_{1}i_{2}\right\rangle \left\langle j_{1}j_{2}\right\vert 
\end{equation}%
depends on $\Delta =$ $\left\langle 11\right\vert \widehat{\rho }%
^{AB}\left\vert 00\right\rangle -\left\langle 01\right\vert \widehat{\rho }%
^{AB}\left\vert 10\right\rangle $ and $\Delta ^{\ast }=$ $\left\langle
00\right\vert \widehat{\rho }^{AB}\left\vert 11\right\rangle -\left\langle
10\right\vert \widehat{\rho }^{AB}\left\vert 01\right\rangle $. The position
of two pairs of matrix elements with $K=2$ is exchanged to construct the
partial transpose. For N subsystems, the set of $K$ distinguishable
subsystems that change state while N-$K$ of the sub-systems do not, can be
chosen in $D_{K}=\frac{N!}{(N-K)!K!}$ distinct ways. However, the number of
matrix elements that are transposed to get $\rho _{K}^{T_{p}}$ from $\rho $
depends on $D_{K}$ and the dimensions $d_{1},d_{2}...d_{N}$ of the
subsystems. For a three qubit system ($ABC$), for example, the two
independent contributions to $\rho _{2}^{T_{A}}(ABC)$ involve the qubit
pairs $AB$ and $AC$, respectively. Consider the three qubit pure state 
\begin{equation}
\Psi ^{ABC}=\sqrt{\mu _{0}}\left\vert 000\right\rangle +\sqrt{\mu _{1}}%
\left( \frac{\left\vert 110\right\rangle +\left\vert 101\right\rangle
+\left\vert 111\right\rangle }{\sqrt{3}}\right) ,\quad \widehat{\rho }%
^{ABC}=\left\vert \Psi ^{ABC}\right\rangle \left\langle \Psi
^{ABC}\right\vert ,  \label{9}
\end{equation}%
with $\mu _{0}+\mu _{1}=1,$ $\sqrt{\mu _{i}}\geq 0$. It is common practice
to trace out subsystem $A$ to obtain the entanglement of $B$ and $C$. State
reduction is an irreversible local operation and it is believed that the
entanglement of the pair $BC$ in the reduced system is either the same or
less than that in the composite system $\widehat{\rho }^{ABC}$. One can,
however, obtain a measure of $2-$way coherences involving a given pair of
subsystems from $2-$way partial transpose constructed by restricting the
transposed matrix elements of $\widehat{\rho }^{ABC}$ to those for which the
state of the third subsystem does not change. For example, $\rho
_{2}^{T_{A-AB}}$ is obtained from the matrix $\rho ^{ABC}$ by applying the
condition%
\begin{eqnarray}
\left\langle i_{1}i_{2}i_{3}\right\vert \widehat{\rho }_{2}^{T_{A-AB}}\left%
\vert j_{1}j_{2}i_{3}\right\rangle  &=&\left\langle
j_{1}i_{2}i_{3}\right\vert \widehat{\rho }\left\vert
i_{1}j_{2}i_{3}\right\rangle ;  \notag \\
\quad if\quad \sum\limits_{m=1}^{3}\left( 1-\delta _{i_{m},j_{m}}\right) 
&=&2,\text{ }\quad   \notag \\
\left\langle i_{1}i_{2}i_{3}\right\vert \widehat{\rho }_{2}^{T_{A-AB}}\left%
\vert j_{1}j_{2}j_{3}\right\rangle  &=&\left\langle
i_{1}i_{2}i_{3}\right\vert \widehat{\rho }\left\vert
j_{1}j_{2}j_{3}\right\rangle ;\quad \text{for all other matrix elements}.
\label{10}
\end{eqnarray}%
Similarly, matrix elements of $\widehat{\rho }_{2}^{T_{A-AC}}$ are related
to matrix elements of the state operator by%
\begin{eqnarray}
\left\langle i_{1}i_{2}i_{3}\right\vert \widehat{\rho }_{2}^{T_{A-AC}}\left%
\vert j_{1}i_{2}j_{3}\right\rangle  &=&\left\langle
j_{1}i_{2}i_{3}\right\vert \widehat{\rho }\left\vert
i_{1}i_{2}j_{3}\right\rangle ;  \notag \\
\quad if\quad \sum\limits_{m=1}^{3}\left( 1-\delta _{i_{m},j_{m}}\right) 
&=&2,\text{ }\quad   \notag \\
\left\langle i_{1}i_{2}i_{3}\right\vert \widehat{\rho }_{2}^{T_{A-AC}}\left%
\vert j_{1}j_{2}j_{3}\right\rangle  &=&\left\langle
i_{1}i_{2}i_{3}\right\vert \widehat{\rho }\left\vert
j_{1}j_{2}j_{3}\right\rangle ;\quad \text{for all other matrix elements}.
\label{10c}
\end{eqnarray}%
The negativities $N_{2}^{A-AB}=\left( \left\Vert \widehat{\rho }%
_{2}^{T_{A-AB}}\right\Vert _{1}-1\right) $ and $N_{2}^{A-AC}=\left(
\left\Vert \widehat{\rho }_{2}^{T_{A-AC}}\right\Vert _{1}-1\right) ,$
measure the $2-$way coherences involving the pairs of subsystems $AB$ and $AC
$, respectively. For the state, $\widehat{\rho }^{ABC}$ of Eq. (\ref{9}) we
get, 
\begin{equation}
N_{2}^{A}=2\sqrt{\frac{2\mu _{0}\mu _{1}}{3}},\qquad N_{2}^{A-AB}=2\sqrt{%
\frac{\mu _{0}\mu _{1}}{3}},  \label{11}
\end{equation}%
$\ $and 
\begin{equation}
N_{2}^{A-AC}=2\sqrt{\frac{\mu _{0}\mu _{1}}{3}}.  \label{12}
\end{equation}%
The squared negativities satisfy the monogamy relation 
\begin{equation}
\left( N_{2}^{A}\right) ^{2}=\left( N_{2}^{A-AC}\right) ^{2}+\left(
N_{2}^{A-AB}\right) ^{2}.  \label{13}
\end{equation}%
If party $C$ measures the state of qubit three and finds it in state $%
\left\vert 0\right\rangle $(this event happens with a probability $P_{0}=%
\frac{2\mu _{0}+1}{3}$) and communicates the result to parties $A$ and $B$,
the entangled state 
\begin{equation}
\Phi _{0}^{AB}=\sqrt{\frac{3\mu _{0}}{2\mu _{0}+1}}\left\vert
00\right\rangle +\sqrt{\frac{\mu _{1}}{2\mu _{0}+1}}\left\vert
11\right\rangle 
\end{equation}%
becomes available to $A$ and $B$. In case the third qubit is found to be in
state $\left\vert 1\right\rangle $ (Probability $P_{1}=\frac{2\mu _{1}}{3}$%
), the state available to $A$ and $B$ is $\Phi _{1}^{AB}=\left( \left\vert
10\right\rangle +\left\vert 11\right\rangle \right) /\sqrt{2},$ a separable
state. The negativity of partial transpose of the state $\left\vert \Phi
_{0}^{AB}\right\rangle \left\langle \Phi _{0}^{AB}\right\vert $ is $%
N^{A}\left( \left\vert \Phi _{0}^{AB}\right\rangle \left\langle \Phi
_{0}^{AB}\right\vert \right) =2\frac{\sqrt{3\mu _{0}\mu _{1}}}{2\mu _{0}+1},$
as such the total free entanglement available to $A$ and $B$ is $%
P_{0}N^{A}\left( \left\vert \Phi _{0}^{AB}\right\rangle \left\langle \Phi
_{0}^{AB}\right\vert \right) =2\sqrt{\frac{\mu _{0}\mu _{1}}{3}}$, which is
the same as $N_{2}^{A-AB}$ obtained from the full state operator. If no
communication between parties takes place, the state tomography by $A$ and $B
$ should find the qubit one and two to be in a mixed state The partially
transposed matrix obtained from 
\begin{equation}
\widehat{\rho }^{AB}(red)=tr_{C}(\widehat{\rho }^{ABC})=P_{0}\left\vert \Phi
_{0}^{AB}\right\rangle \left\langle \Phi _{0}^{AB}\right\vert
+P_{1}\left\vert \Phi _{1}^{AB}\right\rangle \left\langle \Phi
_{1}^{AB}\right\vert ,
\end{equation}%
reads as 
\begin{equation}
\left( \rho ^{AB}(red)\right) ^{T_{A}}=\left[ 
\begin{array}{cccc}
\mu _{0} & 0 & 0 & 0 \\ 
0 & \frac{\mu _{1}}{3} & \sqrt{\frac{\mu _{0}\mu _{1}}{3}} & \frac{\mu _{1}}{%
3} \\ 
0 & \sqrt{\frac{\mu _{0}\mu _{1}}{3}} & 0 & 0 \\ 
0 & \frac{\mu _{1}}{3} & 0 & \frac{2\mu _{1}}{3}%
\end{array}%
\right] .
\end{equation}%
The negativity of partial transpose $\left( \rho ^{AB}(red)\right) ^{T_{A}}$
satisfies 
\begin{equation}
N_{G}^{A}\left( \left( \rho ^{AB}(red)\right) ^{T_{A}}\right) \leq
P_{0}N^{A}\left( \left\vert \Phi _{0}^{AB}\right\rangle \left\langle \Phi
_{0}^{AB}\right\vert \right) +P_{1}N^{A}\left( \left\vert \Phi
_{1}^{AB}\right\rangle \left\langle \Phi _{1}^{AB}\right\vert \right) .
\label{cr}
\end{equation}%
Local unitary operations on subsystems $A$ and $B$ may transform the states $%
\left\vert \Phi _{0}^{AB}\right\rangle $ and $\left\vert \Phi
_{1}^{AB}\right\rangle $ to $\left\vert \chi _{0}^{AB}\right\rangle $ and $%
\left\vert \chi _{1}^{AB}\right\rangle $, respectively with different values
of negativities. Although left hand side of Eq. (\ref{cr}) is invariant
under such operations, the right hand side may be optimized to obtain a
probability distribution that allows the parties A and B to have at their
disposal, at least one state with large bipartite entanglement and high
probability. We notice that a mixed quantum state is very different from a
classical mixed state. No wonder that $N_{G}^{A}\left( \left( \rho
^{AB}(red)\right) ^{T_{A}}\right) ,$ sometimes, fails to detect the optimum
entanglement of subsystem $AB$ which can become available by local
operations and classical communication with $C$. PPT entangled states are a
class of states for which the entanglement detected by global negativity
turns out to be zero.

It is easily verified that for a tripartite system%
\begin{equation}
\widehat{\rho }_{2}^{T_{A}}=\widehat{\rho }_{2}^{T_{A-AB}}+\widehat{\rho }%
_{2}^{T_{A-AC}}-\widehat{\rho }.  \label{14}
\end{equation}%
Generalization to obtain a measure of $K-$way coherences involving a
specific set of $K$ subsystems from the state operator of $N$-partite
composite system is straight forward. No state reduction is involved here.

\section{Contribution of K-way negativity to Global negativity}

Global negativity with respect to a subsystem $p$ can be written as a sum of
partial $K-$way negativities. Using $Tr\left( \widehat{\rho }%
_{G}^{T_{p}}\right) =1,$ the negativity of $\widehat{\rho }_{G}^{T_{p}}$ is
given by 
\begin{equation}
{N}_{G}^{p}=-\frac{2}{d_{p}-1}\sum\limits_{i}\left\langle \Psi
_{i}^{G-}\right\vert \widehat{\rho }_{G}^{T_{p}}\left\vert \Psi
_{i}^{G-}\right\rangle =-\frac{2}{d_{p}-1}\sum\limits_{i}\lambda _{i}^{G-}%
\text{,}  \label{2n}
\end{equation}%
where $\lambda _{i}^{G+}$and $\left\vert \Psi _{i}^{G+}\right\rangle $ ($%
\lambda _{i}^{G-}$and $\left\vert \Psi _{i}^{G-}\right\rangle $) are,
respectively, the positive (negative) eigenvalues and eigenvectors of $%
\widehat{\rho }_{G}^{T_{p}}$. The global transpose with respect to subsystem 
$p$, may also be written as 
\begin{equation}
\widehat{\rho }_{G}^{T_{p}}=\sum\limits_{K=2}^{N}\widehat{\rho }%
_{K}^{T_{p}}-(N-2)\widehat{\rho }.  \label{3n}
\end{equation}%
Substituting Eq. (\ref{3n}) in Eq. (\ref{2n}), and recalling that $\widehat{%
\rho }$ is a positive operator with trace one, we get%
\begin{eqnarray}
-\frac{2}{d_{p}-1}\sum\limits_{i}\lambda _{i}^{G-} &=&-\frac{2}{d_{p}-1}%
\sum\limits_{K=2}^{N}\sum\limits_{i}\left\langle \Psi _{i}^{G-}\right\vert 
\widehat{\rho }_{K}^{T_{p}}\left\vert \Psi _{i}^{G-}\right\rangle  \notag \\
&&+\frac{2(N-2)}{d_{p}-1}\sum\limits_{i}\left\langle \Psi
_{i}^{G-}\right\vert \widehat{\rho }\left\vert \Psi _{i}^{G-}\right\rangle .
\end{eqnarray}%
Defining the partial $K-$way negativity $E_{K}^{p}$ ($K=2$ to $N$) as 
\begin{equation}
E_{K}^{p}=-\frac{2}{d_{p}-1}\sum\limits_{i}\left\langle \Psi
_{i}^{G-}\right\vert \widehat{\rho }_{K}^{T_{p}}\left\vert \Psi
_{i}^{G-}\right\rangle ,  \label{4n}
\end{equation}%
we may split the global negativity for qubit $p$ as%
\begin{equation}
N_{G}^{p}=\sum\limits_{K=2}^{N}E_{K}^{p}-E_{0}^{p},  \label{5n}
\end{equation}%
where%
\begin{equation}
E_{0}^{p}=-\frac{2(N-2)}{d_{p}-1}\sum\limits_{K=2}^{N}\sum\limits_{i}\left%
\langle \Psi _{i}^{G-}\right\vert \widehat{\rho }\left\vert \Psi
_{i}^{G-}\right\rangle .
\end{equation}%
An interesting result is obtained when the global partial transpose has a
single negative eigen value that is 
\begin{equation}
\lambda ^{G-}\left\vert \Psi ^{G-}\right\rangle \left\langle \Psi
^{G-}\right\vert =\sum\limits_{K}\sum\limits_{m}\lambda _{m}^{K-}\left\vert
\Psi _{m}^{K-}\right\rangle \left\langle \Psi _{m}^{K-}\right\vert .
\end{equation}%
In this case $\lambda ^{G-}=\sum\limits_{K}\sum\limits_{m}\lambda _{m}^{K-}$
leading to\bigskip 
\begin{equation}
\left( N_{G}^{p}\right) ^{2}=4\sum\limits_{K=2}^{N}\sum\limits_{m}\left(
\lambda _{m}^{K-}\right) ^{2}.
\end{equation}

For the state $\widehat{\rho }^{ABC}$ of Eq. (\ref{9}), the values 
\begin{equation}
N_{G}^{A}=2\sqrt{\mu _{0}\mu _{1}},\qquad N_{3}^{A}=2\sqrt{\frac{\mu _{0}\mu
_{1}}{3}},\qquad N_{2}^{A}=2\sqrt{\frac{2\mu _{0}\mu _{1}}{3}},
\end{equation}%
are obtained when the global, $2-$way, and $3-$way partial transposes are
taken with respect to first qubit (subsystem $A$). Using the negative eigen
values and eigenvectors of $\widehat{\rho }_{G}^{T_{A}}$ one further gets
the measures of GHZ like entanglement $E_{3}^{A}=\frac{2}{3}\sqrt{\mu
_{0}\mu _{1}}$ and the measure of bipartite entanglement $E_{2}^{A}=\frac{4}{%
3}\sqrt{\mu _{0}\mu _{1}}$. The necessary condition for an $N-$partite pure
state not to have genuine $N-$partite entanglement is that at least one of
the global negativities is zero that is $N_{G}^{p}=0,$ where $p$ is one of
the subsystems or one part of a bipartite split of the composite system.
Recalling that $N_{G}^{p}=\sum\limits_{K=2}^{N}E_{K}^{p}-E_{0}^{p}$ for an $%
N-$partite system in a pure state $\widehat{\rho }$, the separability of
subsystem $p$ implies that $E_{K}^{p}\leq 0,$ or $\widehat{\rho }%
_{K}^{T_{p}}\geqslant 0$, for $K=2$ to $N$. In general, for a system having
only genuine $K-$ partite entanglement, $N_{G}^{p}=0$ for $N-K$ of the
subsystems and $E_{K}^{p}>0$ for at least $K$ subsystems. In addition, the
lowest positive value of the partial non zero $K-$way negativities
determines the $K-$partite entanglement, the same being a collective
property of $K-$subsystems.

\section{Local unitary operations and K-way coherences}

An important point to note is that the trace norm $\left\Vert \widehat{\rho }%
_{K}^{T_{p}}\right\Vert _{1}$ is not invariant under local unitary
rotations, unless, $\left\Vert \widehat{\rho }_{K}^{T_{p}}\right\Vert _{1}=$ 
$\left\Vert \widehat{\rho }_{G}^{T_{p}}\right\Vert _{1}$. \ Any composite
system state $\widehat{\rho }_{2},$ obtained from an N-partite state $%
\widehat{\rho }_{1}$ through entanglement conserving local operations,
differs from the former in being characterized by a different set of $K-$way
Negativities. A canonical state $\widehat{\rho }_{c}$ obtained from $%
\widehat{\rho }$ through entanglement conserving local operations is a state
written in terms of the minimum number of local basis product states \cite%
{cart99}. An N-partite system has N-partite entanglement if for all possible
splits of the system into two subsystems, the global negativity is non zero.
In case N-partite entanglement is generated by N-way coherences, the system
has genuine N-partite entanglement equal to $\min
(E_{N}^{1},E_{N}^{2},...,E_{N}^{N})$, where $%
E_{N}^{1},E_{N}^{2},...,E_{N}^{N}$ are calculated from partial transposes
constructed from the state $\widehat{\rho }_{c}$. We conjecture that for the
state $\widehat{\rho }_{c}$ the quantity $E_{K}^{p}$, as defined in Eq. (\ref%
{4n}) measures the $K-$way entanglement of subsystem $p$ with its
complement. The motivation for using the canonical state stems from the fact
that the coeficients in a canonical form are all local invariants. As such
the calculated contributions $E_{K}^{p}$, being functions of local
invariants having unique values, qualify to be entanglement measures for all
the states lying on the orbit. Consider a three qubit W-like state%
\begin{equation}
\left\vert \Psi _{I}\right\rangle _{123}=\sqrt{a}\left\vert 100\right\rangle
+\sqrt{a}\left\vert 010\right\rangle +\sqrt{1-2a}\left\vert 001\right\rangle
,\quad a\in \left[ \frac{1}{3},\frac{1}{2}\right] 
\end{equation}%
shared by Alice, Bob and Charlie. This state has no $3-$way coherences and
cannot be converted to a GHZ like state by local operations. If two of the
parties, say Alice and Bob get together and perform a CNot gate with qubit
one as control and qubit 2 as target qubit, the transformed state 
\begin{equation}
\left\vert \Psi _{F}\right\rangle _{123}=\sqrt{a}\left\vert 010\right\rangle
+\sqrt{a}\left\vert 110\right\rangle +\sqrt{1-2a}\left\vert 001\right\rangle
,  \label{ghz}
\end{equation}%
is a GHZ like state. What happens to the coherences during this
transformation is manifest in the set of global, 2-way and 3-way
negativities associated with the initial and the final state listed in Table
I. 
\begin{table*}[tbp]
\caption{The global, $2-$way and $3-$way negativities of W-like state $%
\left\vert \Psi _{I}\right\rangle _{123}$ and GHZ like state $\left\vert
\Psi _{F}\right\rangle _{123}$. The measures $E_{2}^{p}$ and $E_{3}^{p}$ ($%
p=1-3$) are also listed.}%
\begin{tabular}{||l||l||l||l||l||l||}
\hline\hline
& $N_{G}^{1}$ & $N_{2}^{1}$ & $N_{3}^{1}$ & $E_{2}^{1}$ & $E_{3}^{1}$ \\ 
\hline\hline
$\left\vert \Psi _{I}\right\rangle _{123}$ & $2\sqrt{a-a^{2}}$ & $2\sqrt{%
a-a^{2}}$ & $0$ & $2\sqrt{a-a^{2}}$ & $0$ \\ \hline\hline
$\left\vert \Psi _{F}\right\rangle _{123}$ & $2\sqrt{a-2a^{2}}$ & $0$ & $2%
\sqrt{a-2a^{2}}$ & $0$ & $2\sqrt{a-2a^{2}}$ \\ \hline\hline
& $N_{G}^{2}$ & $N_{2}^{2}$ & $N_{3}^{2}$ & $E_{2}^{2}$ & $E_{3}^{2}$ \\ 
\hline\hline
$\left\vert \Psi _{I}\right\rangle _{123}$ & $2\sqrt{a-a^{2}}$ & $2\sqrt{%
a-a^{2}}$ & $0$ & $2\sqrt{a-a^{2}}$ & $0$ \\ \hline\hline
$\left\vert \Psi _{F}\right\rangle _{123}$ & $2\sqrt{2a(1-2a)}$ & $2\sqrt{%
a-2a^{2}}$ & $2\sqrt{a-2a^{2}}$ & $\sqrt{2a(1-2a)}$ & $\sqrt{2a(1-2a)}$ \\ 
\hline\hline
& $N_{G}^{3}$ & $N_{2}^{3}$ & $N_{3}^{3}$ & $E_{2}^{3}$ & $E_{3}^{3}$ \\ 
\hline\hline
$\left\vert \Psi _{I}\right\rangle _{123}$ & $2\sqrt{2a(1-2a)}$ & $2\sqrt{%
2a(1-2a)}$ & $0$ & $2\sqrt{2a(1-2a)}$ & $0$ \\ \hline\hline
$\left\vert \Psi _{F}\right\rangle _{123}$ & $2\sqrt{2a(1-2a)}$ & $2\sqrt{%
a-2a^{2}}$ & $2\sqrt{a-2a^{2}}$ & $\sqrt{2a(1-2a)}$ & $\sqrt{2a(1-2a)}$ \\ 
\hline\hline
\end{tabular}%
\end{table*}
All the qubits in state $\left\vert \Psi _{F}\right\rangle _{123}$, have $3-$%
way coherences. In addition qubit two and three also show $2-$way coherence.
The bipartite entanglement of qubit two and three in the the reduced state $%
tr_{1}(\left\vert \Psi _{F}\right\rangle _{123}\left\langle \Psi
_{F}\right\vert )$ is due to coherences $N_{2}^{2}=N_{2}^{3}=2\sqrt{a-2a^{2}}
$. The state reduction destroys genuine tripartite entanglement and increase
the bipartite entanglement.

\section{What\ does the N-way negativity tell about the entanglement
available to N parties sharing the composite system?}

Let $\nu _{N}(\rho )$ be the number of negative eigen values of $\rho
_{N}^{T_{p}-}$ of an entangled state 
\begin{equation}
\rho =\left\vert \Psi \right\rangle \left\langle \Psi \right\vert ,\quad
\,\left\vert \Psi \right\rangle =\sum_{i}\left\vert \Phi _{i}\right\rangle ,
\end{equation}%
where 
\begin{eqnarray}
\left\vert \Phi _{i}\right\rangle &=&a_{i}\left\vert
i_{1}i_{2}...i_{N}\right\rangle +b_{i}\left\vert
j_{1}j_{2}...j_{N}\right\rangle ,\quad i_{m}\neq j_{m},(m=1\text{ to }N), 
\notag \\
\left\langle \Phi _{j}\right\vert \left. \Phi _{i}\right\rangle &=&\sqrt{%
\left\vert a_{i}\right\vert ^{2}+\left\vert b_{i}\right\vert ^{2}}\delta
_{ij},\quad
\end{eqnarray}%
with qubits $1,2,3,...$ held by parties $A,B,C...$ , respectively. A state $%
\left\vert \Psi \right\rangle $ with non increasing $\nu _{N}(\rho )$ under
local unitaries indicates that the state may be transfomed to a state having
minimum number of LBPS by LU on qubit $p$ alone, and has the same negativity
as the canonical state. An N-qubit canonical state may, in turn, be written
as%
\begin{equation}
\left\vert \Psi \right\rangle _{c}=\sum_{i=1}^{\nu _{N}(\rho
_{c})}\left\vert \Phi _{i}\right\rangle +X,\quad \rho _{c}=\left\vert \Psi
\right\rangle _{c\,c}\left\langle \Psi \right\vert ,\quad \left\langle
X\right\vert \left. \Phi _{i}\right\rangle =0\text{ for }i=1-\nu _{N}(\rho
_{c}),
\end{equation}%
such that $\nu _{N}(\rho _{c})$, the number of eigen values of $\left( \rho
_{c}\right) _{N}^{T_{p}-}$ , is non increasing under local unitary
transformations.

In case the negative part of the $N-$way partial transpose 
\begin{equation}
\rho _{N}^{T_{p}-}((\left\vert \Psi \right\rangle \left\langle \Psi
\right\vert ))=-\sum_{i=1}^{\nu }\left\vert a_{i}\right\vert \left\vert
b_{i}\right\vert \left\vert \Psi _{i}^{N-}\right\rangle \left\langle \Psi
_{i}^{N-}\right\vert ,
\end{equation}%
\ lies in a state space orthogonal to that of state $\rho ,$ we get $%
N_{N}^{p}=2\sum\limits_{i}\left\vert a_{i}\right\vert \left\vert
b_{i}\right\vert $ and 
\begin{equation}
\left\vert \Psi _{i}^{N-}\right\rangle =\left( \frac{\left\vert
i_{1}i_{2}...i_{p-1}j_{p}i_{p+1}..i_{N}\right\rangle -\left\vert
j_{1}j_{2}...j_{p-1}i_{p}j_{p+1}...j_{N}\right\rangle }{\sqrt{2}}\right) .
\end{equation}%
For a given state the negativity of $N-$way partial transpose of $\left\vert
\Psi \right\rangle $ with respect to a qubit is the sum of negativities of $%
\nu _{N}$ unnormalized GHZ like states $\left\vert \Phi _{i}\right\rangle $
that may be projected out of the state. In general, the $K-$way negativity
of partial transpose with respect to qubit $p$ depends on the negativities
of $\nu _{K}$ unnormalized GHZ like states of K qubits (where $p$ is one of
the $K$ qubits) that may be projected out of the state. The information
content of a multipartite state is distributed over $\nu =\left(
\sum_{K=2-N}\nu _{K}\right) $ states, with a specific pobability
distribution. The value of $\nu $ is minimized by unitary operations leading
to the canonical state, which is a state written in terms of the lowest
number of LBPS. A canonical state optimizes the probability distribution of
projecting out GHZ-like states. In a canonical state, the\ total information
content of the composite state, is shared by the lowest number of GHZ like
states that may be projected out of state. The motivation for considering
partial $K-$way negativities of the canonical state as entanglement measures
resides in the fact that a canonical state optimizes the form in which the
entanglement is available to a given set of $K-$parties, for all possible
sets of $K-$parties and all possible values of $K$.

\section{Schmidt decomposition and $K-$way negativities}

Consider a set of four parameter two qubit (qubits $A$ and $B$) and one
qutrit ($C$) states%
\begin{equation}
\Psi =a_{0}\left\vert 000\right\rangle +a_{1}\left\vert 101\right\rangle
+a_{2}\left\vert 011\right\rangle +a_{3}\left\vert 112\right\rangle ,\quad
\sum \left\vert a_{i}\right\vert ^{2}=1.  \label{qutrit}
\end{equation}%
The state can be easily written in Schmidt form with respect to qubit A or
qubit B, with Schmidt coefficients, 
\begin{equation}
\mu _{0A}=\sqrt{\left\vert a_{0}\right\vert ^{2}+\left\vert a_{2}\right\vert
^{2}},\qquad \mu _{1A}=\sqrt{\left\vert a_{1}\right\vert ^{2}+\left\vert
a_{3}\right\vert ^{2}},  \label{29}
\end{equation}%
and%
\begin{equation}
\mu _{0B}=\sqrt{\left\vert a_{0}\right\vert ^{2}+\left\vert a_{1}\right\vert
^{2}},\qquad \mu _{1B}=\sqrt{\left\vert a_{2}\right\vert ^{2}+\left\vert
a_{3}\right\vert ^{2}}.  \label{30}
\end{equation}%
\ The calculated negativities 
\begin{eqnarray}
N_{G}^{A} &=&N_{G}^{B}=2\mu _{0A}\mu _{1A},  \notag \\
N_{3}^{A} &=&N_{3}^{B}=2\left\vert a_{0}\right\vert \left\vert
a_{3}\right\vert ,N_{2}^{A}=2\left\vert a_{0}\right\vert \left\vert
a_{1}\right\vert +2\left\vert a_{2}\right\vert \mu _{1A},  \label{31}
\end{eqnarray}%
and%
\begin{equation}
N_{2}^{B}=2\left\vert a_{0}\right\vert \left\vert a_{2}\right\vert
+2\left\vert a_{1}\right\vert \mu _{1B},  \label{32}
\end{equation}%
satisfy the relation%
\begin{equation}
\left( N_{G}^{A}\right) ^{2}\leq \left( \left( N_{2}^{A}\right) ^{2}+\left(
N_{3}^{A}\right) ^{2}\right) ,\quad \left( N_{G}^{B}\right) ^{2}\leq \left(
\left( N_{2}^{B}\right) ^{2}+\left( N_{3}^{B}\right) ^{2}\right) ,
\label{33}
\end{equation}%
and the entanglement measures for the state are 
\begin{equation}
E_{2}^{A}=2\frac{\left\vert a_{0}\right\vert ^{2}\left\vert a_{1}\right\vert
^{2}+\left\vert a_{2}\right\vert ^{2}\mu _{1A}^{2}}{\mu _{0A}\mu _{1A}}%
,\quad E_{3}^{A}=E_{3}^{B}=2\frac{\left\vert a_{0}\right\vert ^{2}\left\vert
a_{3}\right\vert ^{2}}{\mu _{0A}\mu _{1A}},\qquad E_{2}^{B}=2\frac{%
\left\vert a_{0}\right\vert ^{2}\left\vert a_{2}\right\vert ^{2}+\left\vert
a_{1}\right\vert ^{2}\mu _{1B}^{2}}{\mu _{0B}\mu _{1B}}.  \label{34}
\end{equation}%
Rewriting $\widehat{\rho }_{2}^{T_{A}}$ and $\widehat{\rho }_{2}^{T_{B}}$as 
\begin{equation}
\widehat{\rho }_{2}^{T_{A}}=\widehat{\rho }_{2}^{T_{A-AB}}+\widehat{\rho }%
_{2}^{T_{A-AC}}-\widehat{\rho },\quad \widehat{\rho }_{2}^{T_{B}}=\widehat{%
\rho }_{2}^{T_{B-AB}}+\widehat{\rho }_{2}^{T_{B-BC}}-\widehat{\rho },
\end{equation}%
we identify 
\begin{equation}
N_{2}^{A-AB}=2\left\vert a_{1}\right\vert \left\vert a_{2}\right\vert
,N_{2}^{A-AC}=2\left\vert a_{0}\right\vert \left\vert a_{1}\right\vert
+2\left\vert a_{2}\right\vert \left\vert a_{3}\right\vert ,
\end{equation}%
and 
\begin{equation}
N_{2}^{B-BC}=2\left\vert a_{1}\right\vert \left\vert a_{3}\right\vert
+2\left\vert a_{0}\right\vert \left\vert a_{2}\right\vert .
\end{equation}%
The bipartite entanglement measures defined as in Eq. (\ref{5n}) have the
structure $E_{2}^{A}=E_{2}^{AB}+E_{2}^{AC}$ and$\quad
E_{2}^{B}=E_{2}^{AB}+E_{2}^{BC}$, \ where 
\begin{eqnarray}
E_{2}^{AB} &=&2\frac{\left\vert a_{2}\right\vert ^{2}\left\vert
a_{1}\right\vert ^{2}}{\mu _{0A}\mu _{1A}},\qquad E_{2}^{AC}=2\frac{%
\left\vert a_{0}\right\vert ^{2}\left\vert a_{1}\right\vert ^{2}+\left\vert
a_{2}\right\vert ^{2}\left\vert a_{3}\right\vert ^{2}}{\mu _{0A}\mu _{1A}},
\label{35} \\
\qquad E_{2}^{BC} &=&2\frac{\left\vert a_{0}\right\vert ^{2}\left\vert
a_{2}\right\vert ^{2}+\left\vert a_{1}\right\vert ^{2}\left\vert
a_{3}\right\vert ^{2}}{\mu _{0B}\mu _{1B}}.
\end{eqnarray}%
The number of local basis product states in state of Eq. (\ref{qutrit}) can
not be reduced by local rotations. The partial K-way negativities $%
E_{2}^{AB} $ , $E_{2}^{AC}$ and $E_{2}^{BC}$ calculated from pure state
operator determine the bipartite entanglement. The entanglement available to 
$A$ and $B$ if they have no knowledge of $C$, as well as the probabilistic
entanglement of subsystem $AB$ after a measurement has been made by $C$, is
determined by $E_{2}^{AB}$. Whenever, a multipartite state can be written in
Schmidt form for some of the subsystems, analytical expressions for
entanglement measures are easily found. When no analytical expressions are
available, numerical calculations using standard subroutines for calculating
eigenvalues come handy.

To summarize, we have defined global and $K-$way negativities calculated
from global and $K-$way partial transposes, respectively, of an $N-$partite
state operator. For a given partition of a multipartite quantum system,
global negativity measures overall entanglement of parties. The $K-$way
negativities for $2\leq K\leq $ N, on the other hand, provide a measure of $%
K-$way coherences of the system. Entanglement is invariant with respect to
local unitary operations, whereas, coherences are not so. The $K-$way
negativity of partial transpose with respect to a subsystem depends on the
negativities of unnormalized K subsystem GHZ like states available to K
parties. A canonical state is special in that the information contained in
the state is shared by the minimum possible number of K- subsystem GHZ like
states ($2\leq K\leq $ N). Global negativity with respect to a subsystem can
be written as a sum of partial $K-$way negativities. We conjecture that the
partial $K-$way negativities provide measures of $K-$way entanglement for $%
N- $partite canonical states. Analysis of K-way negativities is expected to
provide insight into the entanglement distribution amongst different parts
of a quantum system and point out the direction of entanglement flow during
processes involving time evolution of composite quantum system under unitary
operations. We believe that the use of K-way negativities to characterize
entangled states is an important step towards the understanding of quantum
correlations. As local operations transform $K-$partite entanglement to
entanglement available to $K^{\prime }$ parties, the use of $K-$way
negativities should be helpful in finding the multipartite state that
optimizes the entanglement distribution for implementation of a specific
quantum computation and communication related task. We recall that the
multipartite entanglement is still not well understood, even for
low-dimensional quantum systems. The ideas presented in this article, just
point out a different direction in which one can look for solutions.

Financial support from CNPq, Brazil and Faep-UEL, Brazil is acknowledged.

\end{document}